\documentclass[aps,prb,twocolumn,showpacs,groupedaddress]{revtex4}
\usepackage{amssymb}
\usepackage{graphicx}
\usepackage{bm}
\begin{document}
\title{Quantum phase transition in ultrahigh mobility SiGe/Si/SiGe two-dimensional electron system}
\author{M.~Yu. Melnikov, A.~A. Shashkin, and V.~T. Dolgopolov}
\affiliation{Institute of Solid State Physics, Chernogolovka, Moscow District 142432, Russia}
\author{Amy Y.~X. Zhu and S.~V. Kravchenko}
\affiliation{Physics Department, Northeastern University, Boston, Massachusetts 02115, USA}
\author{S.-H. Huang and C.~W. Liu}
\affiliation{Department of Electrical Engineering and Graduate Institute of Electronics Engineering, National Taiwan University, Taipei 106, Taiwan, and\\ National Nano Device Laboratories, Hsinchu 300, Taiwan}
\begin{abstract}
The metal-insulator transition (MIT) is an exceptional test bed for studying strong electron correlations in two dimensions in the presence of disorder. In the present study, it is found that in contrast to previous experiments on lower-mobility samples, in ultra-high mobility SiGe/Si/SiGe quantum wells the critical electron density, $n_{\text{c}}$, of the MIT becomes smaller than the density, $n_{\text{m}}$, where the effective mass at the Fermi level tends to diverge. Near the topological phase transition expected at $n_{\text{m}}$, the metallic temperature dependence of the resistance should be strengthened, which is consistent with the experimental observation of more than an order of magnitude resistance drop with decreasing temperature below $\sim1$~K.
\end{abstract}
\date{\today}
\maketitle

The zero-magnetic-field metal-insulator transition (MIT) was first observed in a strongly-interacting two-dimensional (2D) electron system in silicon metal-oxide-semiconductor field-effect transistors (MOSFETs) \cite{zavaritskaya1987,kravchenko1994,popovic1997} and subsequently reported in a wide variety of 2D electron and hole systems: $p$-type SiGe heterostructures, GaAs/AlGaAs heterostructures, AlAs heterostructures, ZnO-related heterostructures, \textit{etc}.\ (for a review, see Ref.~\cite{spivak2010}). An important metric defining the MIT is the magnitude of the resistance drop in the metallic regime. The strongest drop of the resistance with decreasing temperature (up to a factor of 7) was reported in Si MOSFETs \cite{kravchenko1994}. In contrast, in spite of much lower level of disorder in GaAs-based structures, the drop in that system has never exceeded a factor of about three. This discrepancy has been attributed primarily to the fact that electrons in Si MOSFETs have two almost degenerate valleys, which further enhances the correlation effects \cite{punnoose2002,punnoose2005}. The importance of these strong interactions in 2D systems has been confirmed recently in the observation of the formation of a quantum electron solid in Si MOSFETs \cite{brussarski2018}.

It has been found that the effective electron mass in Si MOSFET 2D electron systems strongly increases as the electron density is decreased, with a tendency to diverge at a density $n_{\text{m}}$ that lies close to, but is consistently below the critical density, $n_{\text{c}}$, for the MIT (see Refs.~\cite{shashkin2002,mokashi2012}). It has been shown that this mass enhancement is related to the strong metallic temperature dependence of resistance \cite{shashkin2002}. Furthermore, a similar mass increase has been observed in ZnO-related single crystalline heterostructures \cite{kozuka2014}. No distinction has yet been found in these studies between the energy-averaged effective mass, $m$, and the effective mass at the Fermi level, $m_{\text{F}}=p_{\text{F}}/V_{\text{F}}$ (where $p_{\text{F}}$ and $V_{\text{F}}$ are the Fermi momentum and the Fermi velocity). However, it has been shown \cite{melnikov2017} that in ultra-high mobility SiGe/Si/SiGe quantum wells, the behavior of these two values is qualitatively different: while the average mass tends to saturate at very low electron densities, the mass at the Fermi level continues to grow down to the lowest densities at which it can be reliably measured, indicating a band flattening at the Fermi level. In the clean limit reached in the metallic regime \cite{melnikov2017,fleury2010}, one can in principle expect either the presence of a direct relation between the two critical densities $n_{\text{c}}=n_{\text{m}}$ (see, \textit{e.g.}, Ref.~\cite{punnoose2005}) or its absence $n_{\text{c}}<n_{\text{m}}$ (see, \textit{e.g.}, Refs.~\cite{camjayi2008,zverev2012}).

\begin{figure}\vspace{-2mm}
\scalebox{1}{\includegraphics[width=\columnwidth]{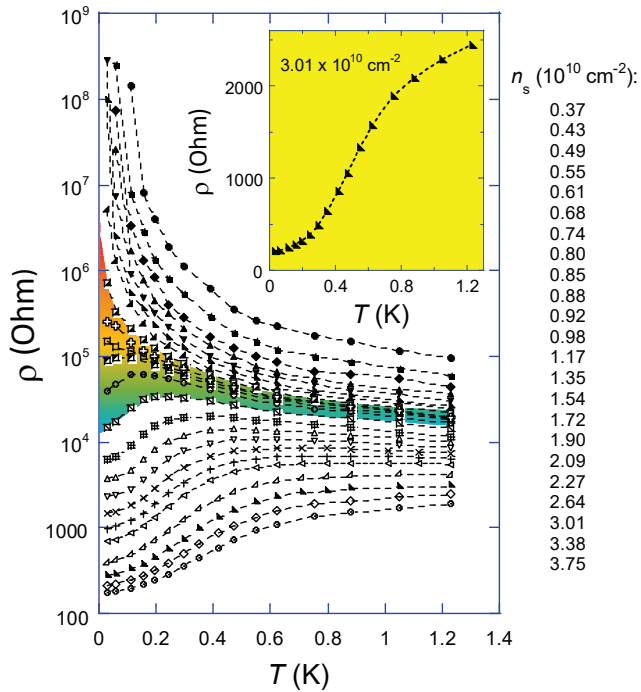}}
\caption{Temperature dependences of the resistivity in a SiGe/Si/SiGe quantum well at different electron densities in zero magnetic field. The inset shows a close-up view of $\rho(T)$ displaying a drop of the resistivity by a factor of 12.}
\label{fig1}
\end{figure}

In this Rapid Communication, we report the study of the metal-insulator transition and the enhanced effective mass at the Fermi level in a strongly-correlated electron system in SiGe/Si/SiGe quantum wells of unprecedentedly high quality. The peak electron mobility in these samples exceeds the peak mobility in the best Si MOSFETs by two orders of magnitude, yet in other respects the two electron systems are similar. In contrast to all previous experiments on low-disordered electron systems, as the residual disorder in an electron system is drastically reduced, we find that the critical electron density of the MIT, $n_{\text{c}}=0.88\pm0.02\times10^{10}$~cm$^{-2}$, determined using three independent methods, becomes smaller than the density where the effective mass at the Fermi level tends to diverge, $n_{\text{m}}=1.1\pm0.1\times10^{10}$~cm$^{-2}$, revealing the qualitative difference between the ultra-low-disorder SiGe/Si/SiGe quantum wells and previously studied electron systems. The finding of $n_{\text{c}}<n_{\text{m}}$ indicates that these two densities are not directly related in the lowest-disorder electron systems, at least. Owing to the difference in the critical electron densities, one expects a topological phase transition at $n_{\text{m}}$, where the Fermi surface breaks into several separate surfaces \cite{zverev2012}. As a result, additional scattering channels appear near $n_{\text{m}}$ on the metallic side of the MIT, which promotes the metallic temperature dependence of the resistance. This is in agreement with the experimental observation of a resistance drop on the metallic side of the transition in our samples by more than an order of magnitude with decreasing temperature from 1.2~K to 30~mK.

Measurements were performed on ultra-high mobility SiGe/Si/SiGe quantum wells similar to those described in Refs.~\cite{melnikov2015,melnikov2017}. The peak electron mobility, $\mu$, in these samples reaches 240~m$^2$/Vs. It is important to note that judging by the appreciably higher quantum electron mobility ($\sim10$~m$^2$/Vs) in the SiGe/Si/SiGe quantum wells compared to that in Si MOSFETs \cite{melnikov2015}, the residual disorder related to both short- and long-range random potential is drastically smaller in the samples used here. The approximately 15~nm wide silicon (001) quantum well is sandwiched between Si$_{0.8}$Ge$_{0.2}$ potential barriers. The samples were patterned in Hall-bar shapes with the distance between the potential probes of 150~$\mu$m and width of 50~$\mu$m using standard photo-lithography. Measurements were carried out in an Oxford TLM-400 dilution refrigerator. Data on the metallic side of the transition were taken by a standard four-terminal lock-in technique in a frequency range 1--10~Hz in the linear response regime. On the insulating side of the transition, the resistance was measured with {\it dc} using a high input impedance electrometer. Since in this regime, the current-voltage ($I-V$) curves are strongly nonlinear, the resistivity was determined from ${\rm d}V/{\rm d}I$ in the linear interval of $I-V$ curves, as $I\rightarrow0$.

The resistivity, $\rho$, as a function of temperature, $T$, in zero magnetic field is shown in Fig.~\ref{fig1} for different electron densities, $n_{\text{s}}$, on both sides of the metal-insulator transition. While at the highest temperature the difference between the resistivities measured at the lowest and highest densities differ by less than two orders of magnitude, at the lowest temperature this difference exceeds six orders of magnitude. Curves near the MIT are indicated by the color-gradated area. We identify the transition point at $n_{\text{c}}=0.88\pm0.02\times10^{10}$~cm$^{-2}$, based on the ${\rm d}\rho/{\rm d}T$ sign-change criterion taking account of the tilted separatrix \cite{punnoose2005}.

We emphasize that the behavior of the electron system under study is qualitatively different from that of the least-disordered Si MOSFETs where the MIT occurs in a strongly-interacting conventional Fermi liquid at $n_{\text{c}}\geq n_{\text{m}}$. The opposite relation $n_{\text{c}}<n_{\text{m}}$ is found in SiGe/Si/SiGe quantum wells so that the MIT occurs in an unconventional Fermi liquid: near the topological phase transition expected at $n_{\text{m}}$, additional scattering channels appear on the metallic side of the MIT, which promotes the metallic temperature dependence of the resistance. This is in agreement with the experimental observation of a low-temperature drop in the resistance by a factor of 12, the highest value reported so far in any 2D system (the inset in Fig.~\ref{fig1}).

Another point of distinction is that the critical density for the MIT is almost an order of magnitude smaller compared to that in the least-disordered Si MOSFETs, where $n_{\text{c}}\approx 8\times10^{10}$~cm$^{-2}$. Such a difference can indeed be expected for an interaction-driven MIT. The interaction parameter, $r_{\text{s}}$, is defined as the ratio of the Coulomb and Fermi energies, $r_{\text{s}}=g_{\text{v}}/(\pi n_{\text{s}})^{1/2}a_{\text{B}}$, where $g_{\text{v}}=2$ is the valley degeneracy and $a_{\text{B}}$ is the effective Bohr radius in the semiconductor. We compare the value of the interaction parameter at the critical density $n_{\text{c}}$ in SiGe/Si/SiGe quantum wells with that in Si MOSFETs (where $r_{\text{s}}\approx20$). The two systems differ by the level of the disorder, the thickness of the 2D layer, and the dielectric constant (7.7 in Si MOSFETs and 12.6 in SiGe/Si/SiGe quantum wells). Due to the higher dielectric constant, the interaction parameter at the same electron density is smaller in SiGe/Si/SiGe quantum wells by approximately 1.6. In addition, the effective $r_{\text{s}}$ value is reduced further due to the much greater thickness of the 2D layer in the SiGe/Si/SiGe quantum wells, which results in a smaller form-factor \cite{ando1982}. Assuming that the effective mass in the SiGe barrier is $\approx0.5\, m_{\text{e}}$ and estimating the barrier height at $\approx25$~meV, we evaluate the penetration of the wave function into the barrier and obtain the effective thickness of the 2D layer $\approx200$~\AA\ compared to $\approx50$~\AA\ in Si MOSFETs. This yields the additional suppression of $r_{\text{s}}$ in the SiGe/Si/SiGe quantum wells compared to Si MOSFETs by a factor of about 1.3. Thus, the electron densities $n_{\text{c}}$ correspond to $r_{\text{s}}\approx20$ in both Si MOSFETs and SiGe/Si/SiGe quantum wells, which is consistent with the results of Ref.~\cite{shashkin07}.

\begin{figure}
\scalebox{1}{\includegraphics[width=\columnwidth]{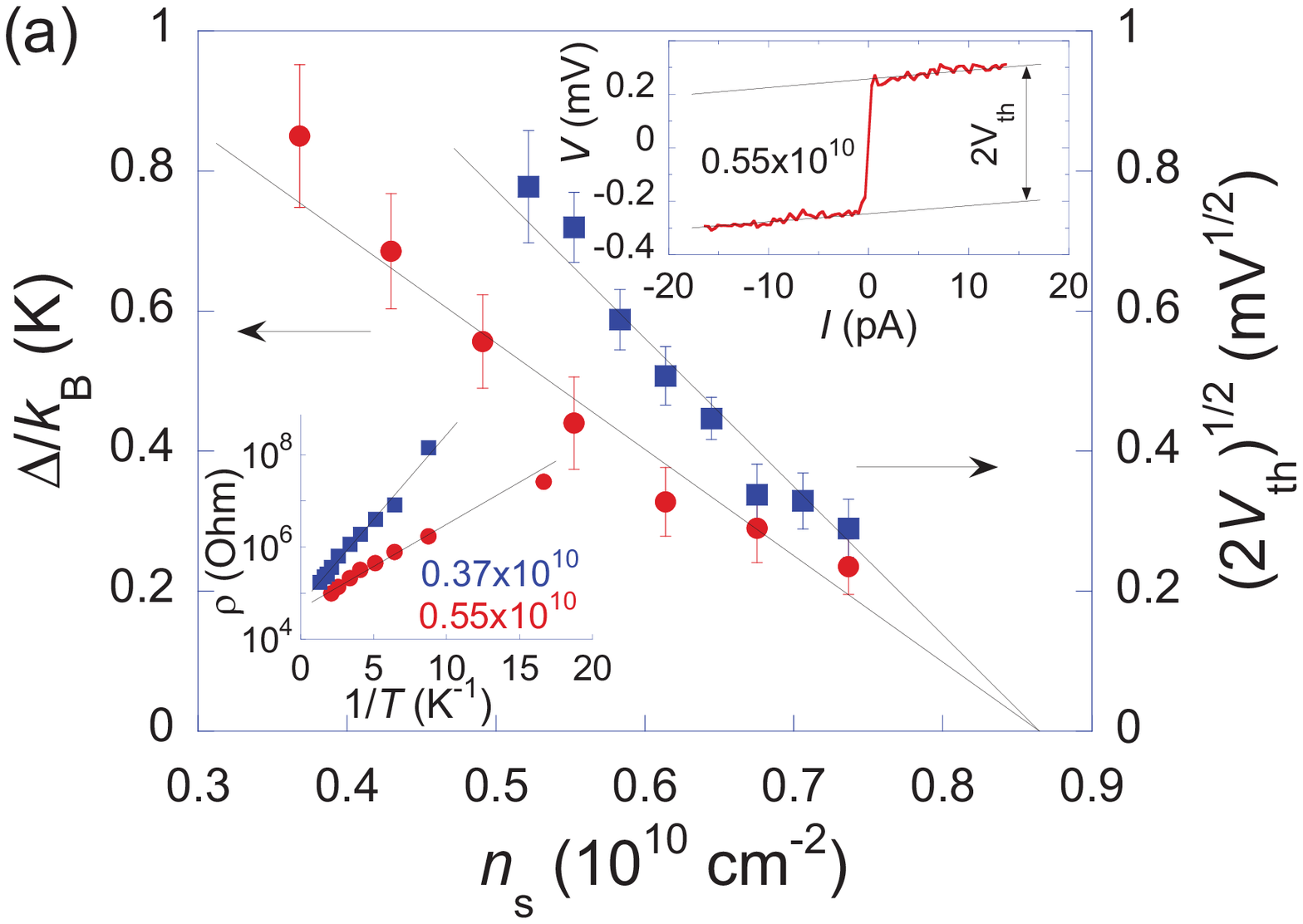}}
\hspace{-5mm}\scalebox{.93}{\includegraphics[width=\columnwidth]{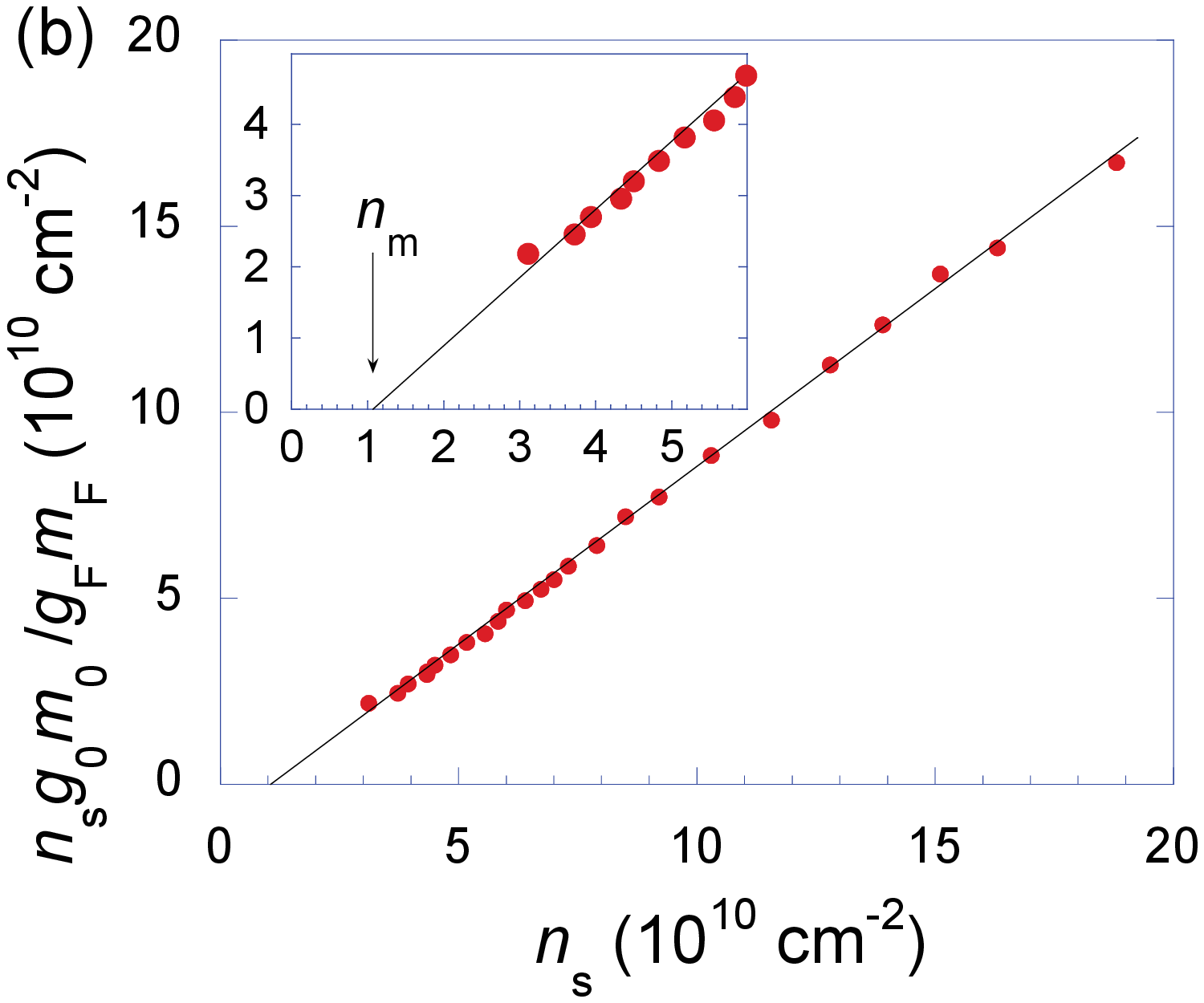}}
\caption{(a)~Activation energy and the square root of the threshold voltage as a function of the electron density in zero magnetic field. Vertical error bars correspond to the experimental uncertainty. The solid lines are linear fits yielding $n_{\text{c}}=0.87\pm0.02\times10^{10}$~cm$^{-2}$. Top inset: Current-voltage characteristic measured at a temperature of 30~mK in zero magnetic field. Bottom inset: Arrhenius plots of the resistivity in the insulating phase for two electron densities. The densities in both insets are indicated in cm$^{-2}$. (b)~Dependence of the effective mass at the Fermi level, $m_{\text{F}}$, on the electron density. The solid line is a linear fit. The experimental uncertainty corresponds to the data dispersion. The inset shows a close-up view of the dependence at low electron densities, where $n_{\text{m}}=1.1\pm0.1\times10^{10}$~cm$^{-2}$.}
\label{fig2}
\end{figure}

The location of the MIT point can also be determined by studying the insulating side of the transition, where the resistance has an activated form, as shown in the bottom inset of Fig.~\ref{fig2}(a); note that the activation energy, $\Delta$, can be determined provided $\Delta>k_{\text{B}}T$. Figure~\ref{fig2}(a) shows the activation energy in temperature units, $\Delta/k_{\text{B}}$, as a function of the electron density (red circles). Near the critical point, this dependence corresponds to the constant thermodynamic density of states and should be linear; the relative accuracy of determination of $\Delta$ increases with increasing activation energy, and the linear fit should be drawn through all data points. The activation energy extrapolates to zero at $n_{\text{c}}=0.87\pm0.02\times10^{10}$~cm$^{-2}$ which coincides, within the experimental uncertainty, with the value of $n_{\text{c}}$ determined from the temperature derivative criterion. Furthermore, in the insulating state, a typical low-temperature $I-V$ curve shows a step-like function: the voltage rises abruptly at low currents and then almost saturates, as seen in the top inset of Fig.~\ref{fig2}(a). The magnitude of the step is $2\, V_{\text{th}}$, where $V_{\text{th}}$ is the threshold voltage. The threshold behavior of the $I-V$ curves has been explained \cite{shashkin1994} within the concept of the breakdown of the insulating phase that occurs when the localized electrons at the Fermi level gain enough energy to reach the mobility edge in an electric field, $V_{\text{th}}/d$, over a distance of the localization length, $L$ (here $d$ is the distance between the potential probes). The values $\Delta/k_{\text{B}}$ and $V_{\text{th}}$ are related via the localization length, which is temperature-independent and diverges near the transition as $L(E_{\text{F}})\propto (E_{\text{c}}-E_{\text{F}})^{-s}$ with exponent $s$ close to unity \cite{shashkin1994} (here $E_{\text{c}}$ is the mobility edge and $E_{\text{F}}$ is the Fermi level). This corresponds to a linear dependence of the square root of $V_{\text{th}}$ on $n_{\text{s}}$ near the MIT, as seen in Fig.~\ref{fig2}(a) (blue squares). The dependence extrapolates to zero at the same electron density as $\Delta/k_{\text{B}}$. A similar analysis has been previously performed \cite{shashkin2001} in a 2D electron system in Si MOSFETs and has yielded similar results, thus adding confidence that the MIT in 2D is a genuine quantum phase transition.

\begin{figure}
\scalebox{1}{\includegraphics[width=\columnwidth]{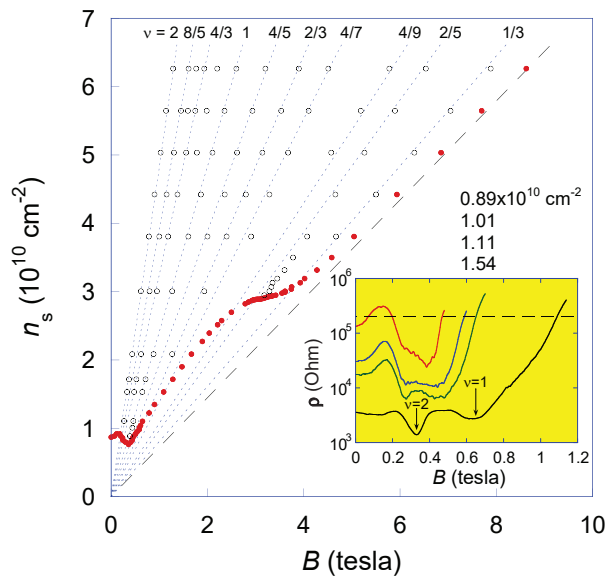}}
\caption{Critical electron density for the MIT as a function of the magnetic field perpendicular to the 2D plane (solid red points). The dashed line indicates the slope of the $n_{\text{c}}(B)$ dependence in the high-field region. Open symbols and dotted lines correspond to the locations of the quantum Hall effect minima in this system. In the inset, $\rho(B)$ dependences used to define the phase boundary are plotted; the cut-off value of the resistivity $\rho=200$~kOhm is indicated by the long-dashed horizontal line; $T=30$~mK.}
\label{fig3}
\end{figure}

The main result of this Rapid Communication paper is shown in Fig.~\ref{fig2}, where we compare the results for $n_{\text{c}}$ to the behavior of the effective electron mass $m_{\text{F}}$ measured at the Fermi level using an analysis of Shubnikov-de Haas oscillations (the procedure of measuring $m_{\text{F}}$ is described in Ref.~\cite{melnikov2017}). In Fig.~\ref{fig2}(b), we plot the product $n_{\text{s}}g_0m_0/g_{\text{F}}m_{\text{F}}$ as a function of the electron density (here $g_0$=2 and $m_0=0.19\, m_{\text{e}}$ are the Land\'e $g$-factor and the effective mass for noninteracting electrons, $m_{\text{e}}$ is the free electron mass, and $g_{\text{F}}\approx g_0$ is the $g$-factor at the Fermi level). The inverse effective mass extrapolates linearly to zero at a density $n_{\text{m}}=1.1\pm0.1\times10^{10}$~cm$^{-2}$, which turns out to be noticeably higher than $n_{\text{c}}$. This finding is in contrast to the results obtained in previous studies on much more disordered electron systems in Si MOSFETs, where a similar change of the inverse effective mass with electron density has been observed but $n_{\text{m}}$ has always been slightly below $n_{\text{c}}$ \cite{shashkin2002,mokashi2012}. We arrive at a conclusion that as the residual disorder in a 2D electron system is decreased, the critical electron density for the MIT becomes lower than the density of the $m_{\text{F}}$ divergence. This indicates that these two densities are not directly related in the lowest-disorder electron systems, at least.

Application of the magnetic field, $B$, perpendicular to the 2D plane affects the critical density of the MIT. Magnetic field dependences of the longitudinal resistivity are shown in the inset to Fig.~\ref{fig3}. The resistivity minimum at the Landau level filling factor $\nu=n_{\text{s}}hc/eB=1$ survives down to electron densities near the MIT. This is similar to the re-entrant behavior that was observed earlier in Si MOSFETs and GaAs/AlGaAs heterostructures \cite{diorio1990,dolgopolov1992,jiang1993,kravchenko1995,qiu2012}. We have chosen the cut-off resistivity for the MIT at $\rho=200$~kOhm, which is close to the value of the critical resistivity for the zero-field MIT at the lowest accessible temperatures; note that the behavior of the $n_{\text{c}}(B)$ phase diagram is only weakly sensitive to the particular cut-off value. Note also that the metallic temperature dependence of the resistance can become insulating with the degree of spin polarization \cite{shashkin2001}, which makes it impossible to use temperature-dependent criteria for the MIT. The resulting phase diagram is shown in Fig.~\ref{fig3}. The critical electron density increases with $B$ at low magnetic fields and then, at the Landau filling factor $\nu=1$, decreases to the value below that for $B=0$. At higher magnetic fields (in the extreme quantum limit), it monotonically grows and exhibits a knee at $\nu=2/5$. Indeed, in this electron system, the longitudinal resistance minimum at $\nu=2/5$ is stronger than that at $\nu=1/3$ (see Ref.~\cite{dolgopolov2018}), in contrast to the strongly-interacting 2D hole system in GaAs/AlGaAs heterostructures \cite{santos1992}. The boundary then continues to grow with a slope corresponding to $\nu\approx0.3$. This is in contrast to the slope of $\nu\approx0.5$ observed in Si MOSFETs in the extreme quantum limit \cite{dolgopolov1992} and interpreted as a consequence of the localization of electrons below half-filling of the lowest Landau level. For comparison, the slope of the high-field boundary in $p$-type GaAs/AlGaAs heterostructures is intermediate (\textit{i.e.}, $\nu\approx0.38$; see Ref.~\cite{qiu2012}). Based on the results obtained in these strongly-interacting carrier systems, we arrive at a conclusion that the critical density for the MIT in the extreme quantum limit is likely to be determined by the level of disorder.

We now discuss the behavior of the critical densities $n_{\text{c}}$ for the $B=0$ MIT and $n_{\text{m}}$ observed in both SiGe/Si/SiGe quantum wells and Si MOSFETs. According to Ref.~\cite{shashkin07}, the effective mass enhancement is independent of disorder, being determined by electron-electron interactions only. The conditions leading to the critical electron density for the $B=0$ MIT are different. Since the value $n_{\text{m}}$ is determined by interactions, the difference between $n_{\text{c}}$ and $n_{\text{m}}$ in SiGe/Si/SiGe quantum wells as compared to Si MOSFETs should be due to $n_{\text{c}}$ being affected by the residual disorder. It is worth noting that according to Ref.~\cite{gold2000} (see also a correction to this paper in Ref.~\cite{dolgopolov2017}), in a moderately-interacting 2D system, the critical density for the $B=0$ MIT should be a power law in the number of impurities: $n_{\text{c}}(\mu)\propto N^{0.75}_{\text{i}}$, which leads to $n_{\text{c}}\propto \mu^{-1.1}$. Therefore, the critical densities should differ by two orders of magnitude in the two systems, which is in contradiction to the experiment. The much weaker change of $n_{\text{c}}$ is likely to reflect the importance of the strong interactions in its behavior.

It follows from the obtained results that the SiGe/Si/SiGe quantum wells are currently a unique electron system with nontrivial topology on the metallic side of the MIT, in which the Fermi surface should break into several separate surfaces at the topological phase transition expected at $n_{\text{m}}$ \cite{zverev2012}. The properties of the electron system can be described qualitatively based on the model of Ref.~\cite{zala01}, where the electron scattering on Friedel oscillations is considered. The resulting linear-in-$T$ correction to conductivity is determined by the slope

\begin{equation}
A=-\frac{(1+\alpha F_0^{\text a})g_{\text F}m_{\text F}}{\pi\hbar^2n_{\text s}},
\label{A}
\end{equation}
where the factor $\alpha$ corresponds to the number of scattering channels, the Fermi liquid parameter $F_0^{\text a}$ is responsible for the renormalization of the $g$-factor $g_{\text F}/g_0=1/(1+F_0^{\text a})$, and $g_{\text F}$ is the $g$-factor at the Fermi level \cite{shashkin2002,zala01}. An increase in the number of scattering channels promotes the metallic temperature dependence of the resistance, in agreement with the experimental observation of the strongest resistance drop with decreasing temperature.

In conclusion, we have studied the metal-insulator transition and the enhanced effective mass at the Fermi level in an ultra-high mobility strongly-interacting 2D electron system in SiGe/Si/SiGe quantum wells. In contrast to previous experiments on low-disordered electron systems, as the residual disorder in an electron system is drastically reduced, we find that the critical electron density $n_{\text{c}}$ of the MIT, obtained using three independent methods, becomes smaller than the density $n_{\text{m}}$ where the effective mass at the Fermi level tends to diverge. Near the topological phase transition expected at $n_{\text{m}}$, additional scattering channels appear on the metallic side of the MIT, which greatly affects the metallic temperature dependence of the resistance. This is consistent with the experimental observation of a resistance drop on the metallic side of the transition by more than an order of magnitude with decreasing temperature below $T\sim1$~K.

We gratefully acknowledge discussions with D. Heiman and V. Kagalovsky. The ISSP group was supported by RFBR 18-02-00368 and 19-02-00196, RAS, and the Russian Ministry of Sciences. The Northeastern group was supported by NSF Grant No. 1309337 and BSF Grant No.\ 2012210. The NTU group was supported by the Ministry of Science and Technology of Taiwan (project nos.\ 106-2622-8-002-001 and 107-2218-E-002-044).

\end{document}